\documentclass[amsmath,amssymb,preprint,showpacs]{revtex4}

\usepackage{graphicx}
\usepackage[usenames]{color}

\begin{document}

\title{Pressure dependence of the optical properties of the charge-density-wave compound LaTe$_2$}
\author{M. Lavagnini, A. Sacchetti and L. Degiorgi} \affiliation{Laboratorium f\"ur
Festk\"orperphysik, ETH - Z\"urich, CH-8093 Z\"urich,
Switzerland}
\author{E. Arcangeletti, L. Baldassarre, P. Postorino and S.
Lupi} \affiliation{CNR-INFM-Coherentia and Dipartimento di Fisica, Universit\`a ``La
Sapienza'', P.le A. Moro 5, I-00185 Rome, Italy}
\author{A. Perucchi} \affiliation{CNR-INFM-Coherentia and Sincrotrone Trieste S.C.p.A., S.S. 14 km 163.5, in Area Science Park, 34012 Basovizza, Trieste, Italy}
\author{K.Y. Shin and I.R. Fisher}
\affiliation{Geballe Laboratory for Advanced Materials and
Department of Applied Physics, Stanford University, Stanford,
California 94305-4045, U.S.A.}

\date{\today}

\begin{abstract}
We report the pressure dependence of the optical response of LaTe$_2$, which is deep in the charge-density-wave (CDW) ground state even at 300 K. The reflectivity spectrum is collected in the mid-infrared spectral range at room temperature and at pressures between 0 and 7 GPa. We extract the energy scale due to the single particle excitation across the CDW gap and the Drude weight. We establish that the gap decreases upon compressing the lattice, while the Drude weight increases. This signals a reduction in the quality of nesting upon applying pressure, therefore inducing a lesser impact of the CDW condensate on the electronic properties of LaTe$_2$. The consequent suppression of the CDW gap leads to a release of additional charge carriers, manifested by the shift of weight from the gap feature into the metallic component of the optical response. On the contrary, the power-law behavior, seen in the optical conductivity at energies above the gap excitation and indicating a weakly interacting limit within the Tomonaga-Luttinger liquid scenario, seems to be only moderately dependent on pressure.
\end{abstract}



\maketitle

\section{Introduction}
Peierls was the first, in the early sixties, to predict the formation of a charge-density-wave (CDW) ground state for one-dimensional (1D) metals, when turning on the electron-phonon interaction \cite{peierls}. The CDW phase transition relates to a balance between electronic energy and lattice structural stability. For 1D metals it is energetically favorable to introduce a lattice distortion which, combined with a so-called Fermi-surface (FS) nesting, leads to a novel collective charge ordering. The formation of the CDW condensate implies the opening of a gap on FS and therefore the lowering of the overall electronic energy \cite{peierls,grunerbook}. Several families of materials, including the transition-metal di- and trichalcogenides, the molybdenum and tungsten oxides such as the blue and purple bronzes, and the organic charge transfer salts, were then discovered after Peierls' pioneering theoretical prediction and intensively studied over several decades \cite{grunerbook,degiorgibook}.

The discovery of superconductivity at high temperature in the low-dimensional layered-like cuprates also induces a revival of interest in the prototype CDW systems, because they provide excellent opportunities for theoretical investigation towards how strongly-correlated electron-phonon systems behave and how electron-phonon interaction affects the band structure. In this context the rare-earth polychalcogenide $R$Te$_n$ ($R$ stays for rare earth, and $n$= 2, 2.5 and 3) systems have attracted a great deal of interest, due to their intrinsic low dimensionality. These materials are also characterized by a layered structure, consisting of corrugated rare-earth-chalcogen slabs alternated with planar chalcogen square lattices (i.e., single layer for di- and double layer for tri-tellurides) \cite{dimasi}. This crystal structure shares some similar features with that of the cuprates, both in terms of the symmetry and also in terms of having two-dimensional layers which are responsible for the electronic properties and which are doped by interleaving block layers. The interest on these systems principally resides in the onset at high temperature of a CDW broken symmetry ground state \cite{dimasi}, driven by a suitable FS nesting. Furthermore, the astonishing discovery of a pressure-induced superconductivity state in CeTe$_2$ below 2.7 K \cite{jung03} competing with a CDW phase (for which the critical temperature $T_{CDW}$ has not yet been identified, but is certainly well above 300 K \cite{dimasi,shin,kang}) and with a rather peculiar magnetic order ($T_N\sim 4.3$ K) \cite{jung00} lately attracted a lot of attention. Such an interplay makes the tellurides an ideal playground in order to investigate the electronic properties with respect to the competition between CDW, magnetic order and superconductivity, and in a broader sense the consequences of the electrons' confinement in the 2D layered-like structure, as well.

Recently, we have intensively investigated the $R$Te$_3$ and $R$Te$_2$ series by optical means \cite{sacchettiprb,sacchettiprl,lavagniniprb}. Optical spectroscopic methods have been generally proven to be a powerful experimental tool in order to address the relevant absorption features associated to the CDW ground state \cite{grunerbook,degiorgibook}. Our first optical reflectivity data on $R$Te$_3$ ($R$=La, Ce, Nd, Sm, Gd, Tb and Dy), collected over an extremely broad spectral range, allowed us to observe both the Drude component and the single-particle peak, ascribed to the contributions due to the free charge carriers and to the excitation across the CDW gap, respectively \cite{sacchettiprb}. We have then measured the pressure dependence of the optical reflectivity on CeTe$_3$ at 300 K (i.e., in the CDW state) \cite{sacchettiprl}. Upon increasing the externally applied pressure the excitation due to the CDW gap decreases, in a quite equivalent manner when compressing the lattice by substituting large with small ionic radius rare-earth elements (i.e., by reducing the lattice constant $a$). Furthermore, the metallic (Drude) weight was found to be moderately enhanced with chemical pressure (i.e., along the rare earth series). These results demonstrate that chemical and applied pressure similarly affect the electronic properties and equivalently govern the onset of the CDW state in $R$Te$_3$. The diminishing impact of the CDW condensate on the FS by reducing the lattice constant is actually the consequence of a quenching of the nesting conditions, driven by the modification of the electronic structure because of the lattice compression \cite{sacchettiprb,sacchettiprl}. 

The latest optical investigation of the related rare-earth di-tellurides $R$Te$_2$ ($R$= La and Ce) confirms our previous findings on $R$Te$_3$ \cite{lavagniniprb}. We have extracted the energy for the CDW gap and found that the CDW collective state gaps a large portion of the Fermi surface. Moreover, it is worth mentioning that for both $R$Te$_2$ and $R$Te$_3$ series we observed a high frequency power-law behavior in the optical conductivity (i.e., at energies larger than the CDW gap) \cite{sacchettiprb,lavagniniprb}. The latter result was found to be compatible with a scenario based on the Tomonaga-Luttinger liquid model for which direct electron-electron interactions and Umklapp processes play a role in the electron dynamics at high energies.

While the single layer $R$Te$_2$ share several common features and similar properties with the related bilayer $R$Te$_3$ materials, there are also important differences. The lattice modulation is somewhat different for each member of the di-tellurides family but is essentially identical for all members of the tri-tellurides family, being characterized in that case by a single unidirectional wave vector. Moreover, $R$Te$_3$ appears to be a line compound, whereas $R$Te$_2$ is known to have a substantial width of formation. As a consequence, while single crystals of $R$Te$_3$ do not exhibit sample-to-sample variation, there is an appreciable difference in the resistivity of single crystals of $R$Te$_2$ even when taken from the same growth batch. Moreover, ARPES and thermodynamic experiments have pointed out the sensitive role played by perturbation, like Te vacancies, on the CDW state of $R$Te$_2$ \cite{shin}. Such a sensitivity has been mainly associated with the observation that the nesting wave vectors, particularly in CeTe$_2$, are somewhat poorly defined \cite{shin}. Therefore, the variation in the Te concentration for different members of the rare-earth di-tellurides makes a systematic study across the whole rare-earth series in $R$Te$_2$ somewhat less meaningful than has been the case for the rare-earth tri-tellurides \cite{sacchettiprb}. Besides the Te deficiencies, changes in the lattice constant may lead to subtle differences in the lattice modulation. Consequently, applied pressure might affect to some extent the electronic structure and the CDW condensate. Analogous to CeTe$_3$ \cite{sacchettiprl} pressure dependent optical investigations may be of great relevance. It is then instructive to establish a comparison between the physical properties of the two classes of rare-earth telluride compounds upon lattice compression. 

In this paper, we present our optical results under externally applied pressure in LaTe$_2$. First, we introduce the investigated material along with the technical details pertaining to the experiment. The data presentation as well as their thorough analysis will be followed by a discussion, primarily focusing the attention on the comparison between the optical properties of the rare-earth polychalcogenides.

\section{Experiment and Results}

LaTe$_2$ single crystals were grown by slow cooling a binary melt so that they are as close to stoichiometry as possible \cite{shin}. Electron microprobe analysis, using elemental standards with an uncertainty of +/- 0.03 in the Te content, determines the precise composition of LaTe$_{1.95}$ for the investigated specimen (which we will indicate as LaTe$_2$ throughout the rest of the paper). Further details about the sample characterization can be found in Ref. \onlinecite{shin}.

A tiny piece of LaTe$_2$ (i.e., approximately $50 \times 50$ $\mu$m$^2$) was cut from the same well characterized specimen previously used in Ref. \onlinecite{lavagniniprb} and was placed on the top surface of a KBr pellet, pre-sintered in the gasket hole of the pressure cell. The gasket was made of stainless steel, 50 $\mu$m
thick under working conditions and with a 200 $\mu$m diameter hole. A clamp-screw diamond
anvil cell (DAC) equipped with high-quality type IIa diamonds (400
$\mu$m culet diameter) was employed for generating high-pressure up to 7 GPa.
Pressure was measured with the standard ruby-fluorescence
technique \cite{mao}. 

Due to the metallic character of the sample,
absorption measurements are not possible on this compound.
Therefore, we carried out optical reflectivity measurements
exploiting the high brilliance of the SISSI infrared beamline at
ELETTRA synchrotron in Trieste \cite{SISSI}. The incident and reflected light
were focused and collected by a cassegrainian-based optical
microscope equipped with a HgCdTe (MCT) detector and coupled to a Bruker
Michelson interferometer equipped with KBr and CaF$_2$ beamsplitter, which allows to explore the 600-11000
cm$^{-1}$ spectral range. In contrast to our previous optical study of CeTe$_3$ under pressure \cite{sacchettiprl}, it was possible to obtain here high quality spectra of the gasket. This has important implications, as discussed below, towards a robust determination of the reference signal and of the correct shape and value of the reflectivity inside the DAC. At each pressure, we therefore measured the
light intensity reflected by the sample $I_S(\omega)$ and by
the (steel) gasket $I_G(\omega)$, obviously at the diamond-specimen interface for both measurements. We thus
obtain the quantity
$R^S_G(\omega)=I_S(\omega)/I_G(\omega)$, so that $I_G(\omega)$ is acting here as reference signal. Measuring the reflected intensity of the gasket at each pressure runs allows us to monitor the variations in the light intensity due to the smooth depletion of the current in the synchrotron storage ring. The final spectra as a function  of pressure were then achieved by multiplying each measured curves by a pressure-independent factor. This latter scaling factor is chosen in such a way so that the final spectra match with the expected $R(\omega)$ at zero pressure inside the cell. The expected $R(\omega)$ at ambient pressure of LaTe$_2$ (inset of Fig. 1) is
calculated from the complex
refractive index at zero pressure \cite{lavagniniprb} and assuming the sample inside the
DAC \cite{wooten,dressel,simulation}. The resulting scaling factor is purely instrumental and corrects possible diffraction effects, induced by non-perfectly flat shape of the sample as well as the underestimation of the reference steel signal. We furthermore emphasize, that the reflectivity $R_G(\omega)=I_G(\omega)/I_{Au}(\omega)$ of steel ($I_{Au}$ being the light intensity reflected by gold) is weakly frequency dependent in the spectral range of interest here. We have checked that the correction of $R_G^{S}(\omega)$ by $R_G$ (i.e., $R_G^{S}*R_G$) does not change the shape of the resulting final spectra, but just renormalized them. In fact, the correction by $R_G$ is already fully encountered by the subsequent rescaling of $R^S_G$ to the expected reflectivity level inside the pressure cell (Fig. 1).

\begin{figure}[!tb]
\center
\includegraphics[width=8.5cm]{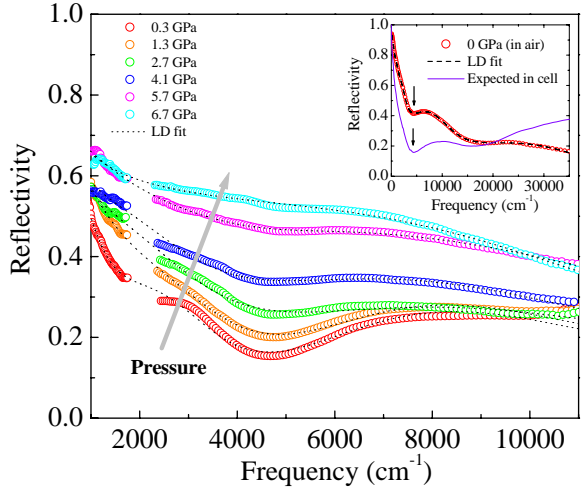}

\caption{(color online) Pressure dependence of $R(\omega)$ in the mid-infrared spectral range of LaTe$_2$ at 300 K. The arrow indicates the trend of the reflectivity data upon increasing pressure. The $R(\omega)$ points in the energy interval of the diamond absorption (i.e., 1700-2300 cm$^{-1}$) have been omitted. The thin dotted lines are fits to the data within the Lorentz-Drude (LD) approach (see text). The inset displays the reflectivity spectrum measured in air from the far infrared up to the ultraviolet together with its LD fit which reproduces in great details the measured $R(\omega)$ at ambient pressure \cite{lavagniniprb}. The expected reflectivity inside the diamond anvil cell at zero pressure \cite{simulation} is reproduced, as well. The arrow is here pointing out the depletion at about 5000 cm$^{-1}$, signaling the onset of the gap absorption.}
\label{Refl}
\end{figure}

Figure 1 displays the pressure dependence of $R(\omega)$ between 10$^3$ and 1.1x10$^4$ cm$^{-1}$, where the strong diamond absorption around 2000 cm$^{-1}$ has been cut out from the spectra, while its inset gives an overall view of $R(\omega)$ at ambient pressure as reproduced from Ref. \onlinecite{lavagniniprb}. It is worth noting that the light spot was precisely limited (by means
of apertures) to the sample area. In contrast to our first pressure-dependent optical investigation on CeTe$_3$ \cite{sacchettiprl}, the $R(\omega)$ spectra of LaTe$_2$ are remarkably smooth and do not display any evidence for interference pattern (between the diamond
windows) because of diffused light. The low pressure $R(\omega)$ reproduces the depletion around 5000 cm$^{-1}$, already observed at ambient pressure (arrow in the inset of Fig. 1) \cite{lavagniniprb} and ascribed to the onset of the CDW gap excitation. The striking feature in Fig. 1 is the progressive increase with increasing pressure of the reflectivity signal, accompanied by the filling-in of the deep minimum in $R(\omega)$ at about 5000
cm$^{-1}$.

\section{Analysis}
In order to extract the optical conductivity
from the pressure-dependent reflectivity spectra, we must
carry out a Kramers-Kronig (KK) analysis. The application of
this method is, however, not trivial because the
measured reflectivity spectra cover a limited frequency range, and the standard KK relation between the
reflectivity and phase needs to be corrected when it is applied to
the sample-diamond interface and the necessary
correction term contains an \textit{a priori} unknown
parameter \cite{pashkin,plaskett}.

For the KK analysis the measured $R(\omega)$ on LaTe$_2$ needs first to be extrapolated to lower and higher frequencies and interpolated within the diamond absorption range
(1700-2300~cm$^{-1}$). Figure 2 highlights the undertaken steps to achieve this goal. First of all, we recall that the complete absorption spectrum of LaTe$_2$ from the far infrared up to the ultraviolet at ambient pressure can be well reproduced within the Lorentz-Drude approach \cite{lavagniniprb}. It consists in fitting the dielectric function by the following expression \cite{wooten,dressel}:

\begin{eqnarray}
\nonumber \tilde{\epsilon}(\omega) & = & \epsilon_1(\omega)
+i\epsilon_2(\omega) =
\\ & = & \epsilon_{\infty}-\frac{\omega_p^2}{\omega^2+i \omega
\gamma_D}+\sum_j \frac{S_j^2}{\omega^2-\omega_j^2-i \omega
\gamma_j},
\end{eqnarray}
where $\epsilon_{\infty}$ is the optical dielectric constant,
$\omega_p$ and $\gamma_D$ are the plasma frequency and the width
of the Drude peak, whereas $\omega_j$, $\gamma_j$, and $S^2_j$ are
the center-peak frequency, the width, and the mode strength for
the $j$-th Lorentz harmonic oscillator (h.o.), respectively. The knowledge of $\tilde{\epsilon}(\omega)$ gives us access to all optical functions and finally allows us to reproduce the measured $R(\omega)$ spectra. The optical properties of $R$Te$_2$ at ambient pressure are well described by one Drude term for the metallic component and three Lorentz h.o.'s accounting for the broad mid-infrared feature, then ascribed to the single particle peak excitation. One additional h.o. is also considered in the fit procedure in order to mimic the onset of the electronic interband transitions (see inset of Fig. 2 in Ref. \onlinecite{lavagniniprb}). The fit quality over the entire spectral range is remarkably good, as shown in the inset of Fig. 1 for our LaTe$_2$ sample \cite{lavagniniprb}.

\begin{figure}[!tb]
\center
\includegraphics[width=8.5cm]{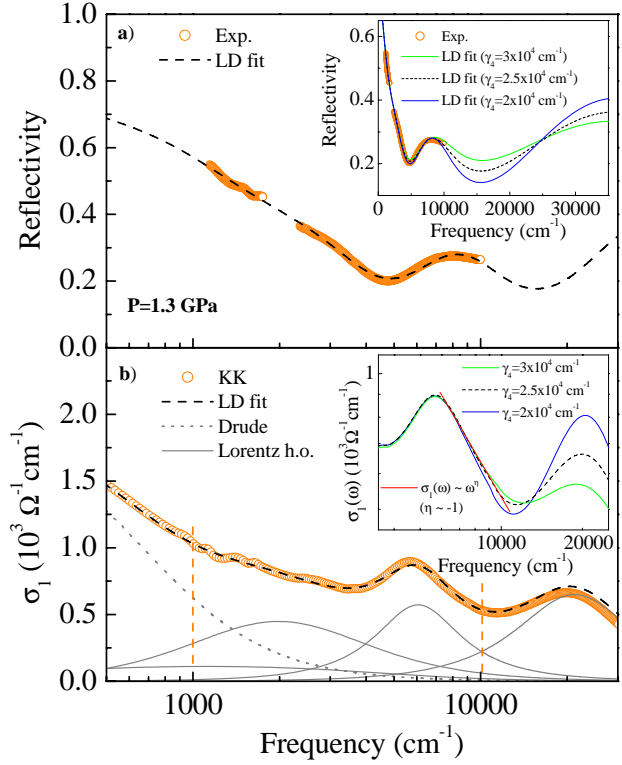}

\caption{(color online) (a) Measured $R(\omega)$ of LaTe$_2$ at 1.3 GPa and its extension based on the Lorentz-Drude (LD) fit (see text). (b) Real part $\sigma_1(\omega)$ of the complex optical conductivity achieved through Kramers-Kronig (KK) transformation of the spectrum in panel (a) and its reproduction within the Lorentz-Drude fit. The fit components are displayed, as well. The dashed vertical lines in panel (b) highlight the spectral range, where the original $R(\omega)$ data were collected. The inset in panel (a) shows three different extrapolations of $R(\omega)$ at high frequencies (see text), while the inset in panel (b) reproduces the corresponding $\sigma_1(\omega)$, plotted on a bi-logarithmic scale. The power-law behavior ($\sigma_1(\omega)\sim \omega^{\eta}$) at frequencies above 6000 cm$^{-1}$ is unaffected by the various extrapolations and $\eta\sim$ -1 with a variation of about $\pm$0.1 for the three extrapolations.}
\label{KK_procedure}
\end{figure}

Using the same number of fit components as at ambient pressure \cite{lavagniniprb}, we can also reproduce the $R(\omega)$ spectra under pressure. We directly fit $R(\omega)$ in the measured spectral range (Fig. 1), using eq. (1) and obviously accounting for the sample inside the DAC \cite{simulation}. Even though we address a limited energy interval, we are able to achieve a rather precise fit of $R(\omega)$ as a function of pressure (thin dotted lines in Fig. 1). This phenomenological approach also enables us to extrapolate the $R(\omega)$ spectra beyond the experimentally available energy range and to even interpolate $R(\omega)$ in the energy region of the diamond absorption. Such an extrapolation at lower and higher energies is shown, as an example, for the data at 1.3 GPa in Fig. 2a. The related optical conductivity $\sigma_1(\omega)$ at 1.3 GPa, calculated within the Lorentz-Drude fit, is displayed in Fig. 2b, along with its own fit components.

The precise shape of $\sigma_1(\omega)$ and its peculiar frequency dependence, including the possible power-law behavior at high frequencies discussed below, are sensitively governed by subtle changes of the measured $R(\omega)$. The Lorentz-Drude reconstruction of $\sigma_1(\omega)$ is then not enough, since $\sigma_1(\omega)$ calculated within this phenomenological method might suffer to some extent from the constraints, imposed by the use of Lorentz h.o.'s. We therefore perform reliable KK transformations, following the procedure successfully employed by Pashkin \emph{et al.} for the organic Bechgaard salt (TMTTF)$_2$AsF$_6$ \cite{pashkin}. The KK
relation for the phase $\phi$ of the reflectivity $R(\omega)$ has the
following form \cite{plaskett,mcdonald}:
\begin{equation}
  \phi(\omega_0)=-\frac{\omega_0}{\pi}P \int_{0}^{+\infty}\frac{\ln R(\omega)}{\omega^2-\omega_0^2}d\omega+\left[\pi-2\arctan\frac{\omega_\beta}{\omega_0}\right],
  \label{eq:KK}
\end{equation}
where $\omega_\beta$ is the position of the reflectivity pole on
the imaginary axis in the complex frequency plane. In case of
measurements on the sample-air interface, $\omega_\beta$ tends
towards infinity and the second term vanishes. For the
sample-diamond interface the second term must, however, be taken into
account. The criterium for the proper value of $\omega_\beta$ is
the agreement between the optical conductivity obtained by the KK
analysis and that from the initial fit \cite{pashkin}.

Figure 2b well illustrates the self-consistency of the applied data analysis for the 1.3 GPa data. The comparison between the $\sigma_1(\omega)$ spectra, obtained first through KK transformation of the extended $R(\omega)$ data of Fig. 2a and second from the direct Lorentz-Drude fit, is indeed astonishingly good and well emphasizes the reliability of this procedure. Table I summarizes for all studied pressures the $\omega_{\beta}$ values, which allow the best agreement between the Lorentz-Drude calculation of $\sigma_1(\omega)$ and the output of the KK transformations.

\begin{table}[!t]
\centering
\begin{tabular*}{\columnwidth}{@{\extracolsep{\fill}}cccccc}
\\\hline\hline
P(GPa) & $\omega_{\beta}(cm^{-1}) $ & $\omega_{SP}(cm^{-1})$ & $\omega_{p}(cm^{-1})$ & $\Phi$
& $\eta$ \\
\hline
0.3 & 8000 & 4754 & 7656 & 0.16 & -1.1 \\
1.3 & 8500 & 4206 & 8953 & 0.21 & -1.0 \\
2.7 & 9600 & 4316 & 9053 & 0.16 & -0.7 \\
4.1 & 9800 & 3528 & 10510 & 0.22 & -0.8 \\
5.7 & 12350 & 2425 & 14000 & 0.26 & -1.0 \\
6.7 & 11500 & 2285 & 15000 & 0.36 & -1.0 \\

\hline\hline
\end{tabular*}

\caption{Pressure dependence of the reflectivity energy pole $\omega_{\beta}$, the single particle peak $\omega_{SP}$, the plasma frequency $\omega_p$, the fraction $\Phi$ of the ungapped Fermi surface and the power-law exponent $\eta$.} \label{Tab}
\end{table}

We also took great care to check the effect of the high frequency extrapolations of the measured $R(\omega)$; a rather sensitive issue when applying the KK analysis. The inset of Fig. 2a shows $R(\omega)$ at 1.3 GPa with three different extrapolations, which have been \emph{ad hoc} manipulated by changing the width of the fourth h.o. at $\omega_4\sim$ 1.2x10$^4$ cm$^{-1}$. The first extrapolation considers $\gamma_4$= 2.5x10$^4$ cm$^{-1}$ for the fourth h.o., which also corresponds to the best fit of the measured $R(\omega)$ (dashed line in the main panel of Fig. 2a). The other two extrapolations were obtained with $\gamma_4$= 2x10$^4$ and 3x10$^4$ cm$^{-1}$, respectively \cite{comment}. The resulting altered $R(\omega)$ at high frequencies smoothly joins the rest of the (measured) $R(\omega)$ signal at about 10$^4$ cm$^{-1}$. The inset of Fig. 2b compares the real part $\sigma_1(\omega)$ of the complex optical conductivity obtained by the KK transformations of the $R(\omega)$ spectra with the three different extrapolations (inset of Fig. 2a). Due to the rather local character of the KK transformations, the final result is less affected by the extrapolations of $R(\omega)$ at frequencies above 10$^4$ cm$^{-1}$ and their impact on the frequency dependence of $\sigma_1(\omega)$ below 10$^4$ cm$^{-1}$ is very moderate. The $\sigma_1(\omega)$ spectra are almost identical and start to deviate from each other at the very upper end of the measured spectral range. This also means that we can trust our data and the corresponding KK analysis all the way up to the high frequency limit of about $\omega\sim$1.1x10$^4$ cm$^{-1}$, reached in our experiment. We have also considered other (more crude) routes to artificially extrapolate $R(\omega)$ at high frequencies (e.g., by simply multiplying the spectra above 10$^4$ cm$^{-1}$ by a factor), reaching however similar conclusions.

\begin{figure}[!tb]
\center
\includegraphics[width=8.5cm]{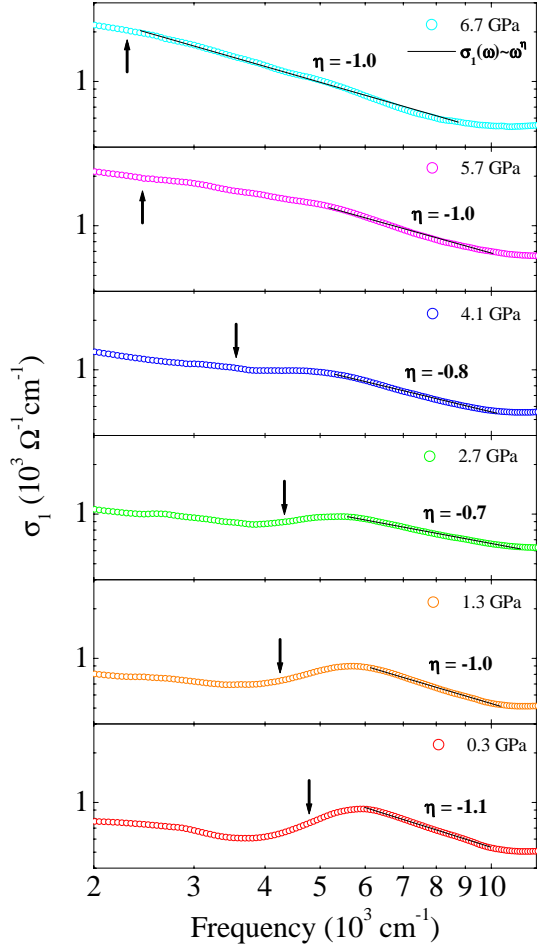}

\caption{(color online) Real part $\sigma_1(\omega)$ of the complex optical conductivity of LaTe$_2$ at various pressures, plotted on a bi-logarithmic scale. The arrows indicate the position of $\omega_{SP}$ (eq. (3)). The power-law behavior ($\sigma_1(\omega)\sim \omega^{\eta}$) is also displayed with the resulting exponent $\eta$.}
\label{sigma}
\end{figure}

\section{Discussion}
The optical conductivity at all measured pressures is shown in Fig. 3 in the spectral range 2x10$^3$ $\leq \omega \leq$ 1.2x10$^4$ cm$^{-1}$, which essentially lies within the energy interval (vertical dashed lines in Fig. 2b) covered by the measurement of $R(\omega)$ under pressure. Particularly at low pressures, two main features are immediately well recognized: the low frequency spectral weight and the mid-infrared absorption, peaked at about 6000 cm$^{-1}$. Indeed, the finite optical conductivity below 3000 cm$^{-1}$ signals the onset of an effective metallic contribution, which is associated to the Drude response. According to the arguments and experimental evidences presented in Ref. \onlinecite{lavagniniprb}, we ascribe the mid-infrared absorption to the single particle peak excitation across the CDW gap. The application of pressure causes a gradual shift of the mid-infrared feature to lower frequencies in $\sigma_1(\omega)$, so that it progressively merges into the high frequency tail of the metallic (Drude) response. This pairs with the already anticipated behavior of $R(\omega)$ (Fig. 1), where the depletion at about 5000 cm$^{-1}$ disappears with increasing pressure. The trend in $\sigma_1(\omega)$ bears a striking similarity with what has been recognized in previous data about the chemical and applied pressure dependence of the optical properties in the related $R$Te$_3$ compounds \cite{sacchettiprb,sacchettiprl}.

It is instructive to extract the relevant energy scales shaping the absorption spectrum: the single particle peak frequency $\omega_{SP}$, due to the excitations across the CDW gap, and the Drude plasma frequency $\omega_p$. To this end, we exploit the output of the Lorentz-Drude fit to our spectra, described above. In our previous investigation of the pressure dependence of $R(\omega)$ in CeTe$_3$ \cite{sacchettiprl}, we failed to collect data below the diamond absorption at about 2000 cm$^{-1}$. In LaTe$_2$ on the contrary, we could extend the measurable spectral range of $R(\omega)$ even in the interval between 1000 and 1700 cm$^{-1}$. This helps in order to better pin down the fit of the Drude component to the measured data set, allowing this way a more realistic estimation of $\omega_p$ under pressure. Consequently, the Lorentz-Drude fit permits us to establish, how the spectral weight (which is proportional to $\omega_p^2$ for the Drude term and to $S_j^2$ for the Lorentz h.o.) is distributed among the various components (Fig. 2b and 3). The plasma frequency $\omega_p$ is listed in Table I.

Analogous to our recent optical investigations on $R$Te$_2$ and $R$Te$_3$ \cite{sacchettiprb,sacchettiprl,lavagniniprb}, we then define the average weighted energy $\omega_{SP}$ as follows:
\begin{equation}
\omega_{SP}=\frac{\sum_{j=1}^3 \omega_j S_j^2}{\sum_{j=1}^3
S_j^2}.
\end{equation}
The sum is over the first three h.o.'s. $\omega_{SP}$ is reported in Table I \cite{comment2} and its position is also indicated by the arrows in Fig. 3. The decrease of $\omega_{SP}$ with pressure confirms the qualitative observation, made above, about the merging of the mid-infrared feature into the Drude component of $\sigma_1(\omega)$. It is worthwhile to compare the pressure dependence of $\omega_{SP}$ for LaTe$_2$ with that of CeTe$_3$ \cite{sacchettiprl}. Such a comparison is shown in Fig. 4, which also displays $\omega_{SP}$ for the rare-earth tri-telluride series (i.e., chemical pressure) \cite{sacchettiprb}. To allow this comparison we plot $\omega_{SP}$ as a function of the lattice constant $a$. We note that since the pressure dependence of the lattice parameters of LaTe$_2$ is not known, we must rely on the crude but effective approach based on the Murnaghan equation \cite{murnaghan,a(p)}, already applied for CeTe$_3$ \cite{sacchettiprl}. It is quite evident that by compressing the lattice there is a progressive reduction of $\omega_{SP}$ \cite{sacchettiprb,sacchettiprl}. The close similarity in the pressure dependence of the optical properties of $R$Te$_2$ and $R$Te$_3$ indicates that pressure affects the electronic structure of these two sets of compounds similarly, irrespective of whether the materials contain single or double Te layers. Consequently, the striking variation in $T_{CDW}$ across the rare earth series for $R$Te$_3$ \cite{ru} cannot be attributed primarily to variation in the bilayer splitting of the FS as a consequence of chemical pressure. In that case one would anticipate a more dramatic variation with respect to applied pressure for the $R$Te$_3$ compounds than for $R$Te$_2$, which is not observed. 

The suppression of $\omega_{SP}$ upon reducing the lattice constant is also accompanied by an enhancement of the plasma frequency $\omega_{p}$. This is shown in the inset of Fig. 4, noting that the linear interpolation is here meant as guide to the eyes, while pressure is obviously an implicit variable. These findings emphasize that pressure affects the electronic structure in such a way as to reduce the impact of the CDW condensate. The decrease of $\omega_{SP}$ upon compressing the lattice concomitantly occurs with the release of itinerant charge carriers (inset of Fig. 4), which therefore increases the effective Drude weight (i.e., the plasma frequency). 

\begin{figure}[!tb]
\center
\includegraphics[width=8.5cm]{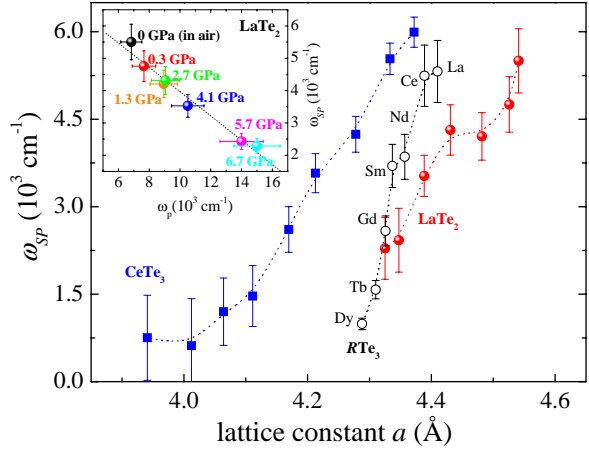}
\caption{(color online) Single particle peak energy $\omega_{SP}$ as a function of the lattice constant $a$ for LaTe$_2$, CeTe$_3$ \cite{sacchettiprl} and the $R$Te$_3$ series \cite{sacchettiprb}. Inset: single particle peak energy $\omega_{SP}$ versus plasma frequency $\omega_p$ for LaTe$_2$, as a function of pressure. Pressure is here an implicit variable.}
\label{gap-vs-a}
\end{figure}

Further support to the above considerations comes from sum rule arguments. Following indeed the well established concept about the spectral weight distribution, introduced in our previous work on NbSe$_3$ \cite{perucchi} and then successfully applied on $R$Te$_2$ and $R$Te$_3$ as well \cite{sacchettiprb,lavagniniprb}, we can define the ratio:
\begin{equation}
\Phi=\omega_p^2/(\omega_p^2+\sum_{j=1}^3 S_j^2) 
\end{equation}
between the Drude weight in the CDW state and the total spectral weight of the hypothetical normal state. This latter quantity is achieved by assuming that above $T_{CDW}$ the weight of the single particle peak (i.e., $\sum_{j=1}^3 S_j^2$) merges together with the Drude weight. Equation (4) tells us how much of FS survives in the CDW state and is not gapped by the formation of the CDW condensate. $\Phi$ is displayed in Table I and turns out to generally increase  upon compressing the lattice. This increase of $\Phi$ with pressure is pretty much in trend with results on $R$Te$_2$ \cite{lavagniniprb} and $R$Te$_3$ \cite{sacchettiprb} and emphasizes the lesser impact of the CDW on FS upon reducing the lattice constant.

The extension, achieved here, of the measurable spectral range up to 1.1x10$^4$ cm$^{-1}$ is of great relevance for the assessment of the pressure dependence of the power-law behavior in $\sigma_1(\omega)$ at $\omega\ge\omega_{SP}$. Figure 3 reports the best power-law $\sigma_1(\omega)\sim \omega^{\eta}$ of LaTe$_2$ under pressure, in the spectral range above the CDW gap absorption. The exponents $\eta$ are summarized in Table I, with an estimation-error of about $\pm$ 0.1. The values of $\eta$ are very close to -1, which compare fairly well with the previous results on LaTe$_2$ and CeTe$_2$ \cite{lavagniniprb}, as well as on the $R$Te$_3$ series at ambient pressure \cite{sacchettiprb}. Even though we have seek the largest energy interval, for which such a power-law in $\sigma_1(\omega)$ applies, in most cases though, it is appropriate for an energy range extending over less than a decade. This means that caution should be placed on the power-law behavior given the rather small frequency interval over which it is extracted. Nonetheless, the power-law in $\sigma_1(\omega)$ is found within the spectral range originally covered by the $R(\omega)$ measurements and it is independent from any extrapolation effects of $R(\omega)$. This is clearly demonstrated in the inset of Fig. 2b, where the power-law behaviors in $\sigma_1(\omega)$ for the three different extrapolations of $R(\omega)$ (inset of Fig. 2a) are in fact almost identical, giving the exponent $\eta \sim$ -1.0 $\pm$0.1.

Power-law behaviors are expected in a wealth of physical quantities, when confining electron in low dimensions and considering the effects of interactions. This occurs particularly in one dimension, because the direct electron-electron interaction is indeed unavoidable and the quasiparticle concept breaks down. The appropriate theoretical framework is based on the Tomonaga-Luttinger liquid scenario, for which the exponent $\eta$ of the power-law is a measurement of the strength of interactions \cite{giamarchi}. The clearest optical evidence for a Tomonaga-Luttinger liquid behavior has been achieved so far in the prototype one-dimensional organic Bechgaard salts \cite{vescoliscience}. A similar power-law decay of $\sigma_1(\omega)$, as seen in the organics at energies larger than the gap, is also predicted for the CDW state. If the Tomonaga-Luttinger liquid scenario is applicable to the rare-earth tellurides, it would give evidence for direct interaction between electrons, as source of Umklapp scattering \cite{giamarchi}, and would imply the non-negligible contribution of 1D correlation effects in the physics of these low-dimensional systems \cite{giamarchi}. Moreover, values of $\eta$ of about -1 might indicate that the rare-earth di-tellurides are close to the weakly interacting limit within the Tomonaga-Luttinger liquid framework \cite{commentpl}. The most puzzling finding, however, is the rather negligible pressure dependence of $\eta$. This would suggest that (1D) correlation effects do not change dramatically upon compressing the lattice. For the range of pressures considered, we might even speculate that compressing the lattice does not induce any remarkable dimensionality crossover.

\section{Conclusions}
We have reported the optical response of LaTe$_2$ under externally applied pressure, focusing our attention on the single particle peak excitation across the CDW gap. We were able to cover a rather large spectral range in the mid-infrared and to reach high pressures up to 7 GPa. These are important prerequisites in order to establish the progressive closing of the CDW gap and the enhancement of the Drude weight upon compressing the lattice. The investigated spectral range is broad enough to allow furthermore educated guesses on the issue to which extent the applied pressure may influence the effect of electron-electron interactions. It turns out that lattice compression moderately affects the electronic correlations in LaTe$_2$.

\begin{acknowledgments}
The authors wish to thank J. M\"uller for technical help and T. Giamarchi for fruitful discussions. One of us (A.S.) wishes to acknowledge financial support from the Della Riccia Foundation. This work has been supported
by the Swiss National Foundation for the Scientific Research
within the NCCR MaNEP pool. This work is also supported by the
Department of Energy, Office of Basic Energy Sciences under
contract DE-AC02-76SF00515.
\end{acknowledgments}


\begin{thebibliography}{99}
\bibitem{comment} The change of $\gamma_4$ implies a minor adjustment of the parameters of $S_{3}^2$ and $\gamma_3$ (strength and damping) of the third Lorentz h.o., leaving however the other ones unchanged. Apart from the obvious changes of $R(\omega)$ above the upper limit of our measurement, the fit quality of $R(\omega)$ remains outstanding in the measured spectral range.
\bibitem{comment2} The single particle peak frequency $\omega_{SP}$ (eq. (3)) does not change significantly as consequence of the three extrapolations of $R(\omega)$ (inset of Fig. 2a) \cite{comment}. The same applies for the plasma frequencies $\omega_p$, which is totally unaffected by the shape of $R(\omega)$ at high frequencies.
\bibitem{a(p)} From the phononic term $\beta T^3$ in the specific heat ($\beta$=8x10$^{-4}$ J/molK$^4$  \cite{shin}) we extract the bulk modulus $B_0$=$\rho v_s^2$=29.3 GPa, $v_s$ being the sound velocity. Assuming the linear pressure dependence of $B(P)$ it follows that the pressure dependence of the volume $V(P)$ is achieved through: $V(P)=V(0)\Big(1+\frac{B'}{B_0}P\Big)^{-1/B'}$ \cite{murnaghan}, with $B'$ varying between 4 and 8. The pressure dependence of the lattice constant is then $a(P)=a(0)[V(P)/V(0)]^{1/3}$, considered as the average between the estimation of $V(p)$ for $B'$=4 and 8 \cite{sacchettiprl}.
\bibitem{commentpl} We note that the exponent $\eta$ for a one-dimensional band insulator would be totally different, ranging between -2, when coupling to phonons is included, and -3, when the lattice is assumed to be rigid and only umklapp scattering off the single-period lattice potential is possible $\lbrack$V. Vescoli \textit{et al.}, \textit{Phys. Rev. Lett.}
\textbf{84}, 1272 (2000)$\rbrack$. Our exponents $\eta$ are moreover different from the expectation in the case of the decay of $\sigma_1(\omega)$ due to the high frequency tail of a single Lorentzian, for which $\eta$ is equal to -2.
\end{thebibliography}
\end{document}